\documentclass[twocolumn,aps,floatfix,showpacs,tightenlines,superscriptaddress,amsmath,amssymb]{revtex4}
\pdfoutput=1
\usepackage{amsthm,amsbsy,epsfig,color,graphicx,times,bbm,enumitem}
\usepackage[ansinew]{inputenc}
\usepackage[english]{babel}
\usepackage{accents}
\usepackage{braket}
\usepackage{amsfonts}
\usepackage{booktabs}
\usepackage{tabularx}
\usepackage{array}
\usepackage{hyperref}
\hypersetup{pdfstartview=FitH}
\usepackage{latexsym}
\usepackage{multirow}
\usepackage{longtable}

\begin{document}

\title{The Interplay between Frustration and Entanglement in Many-Body Systems}

\author{S. M. Giampaolo}
\affiliation{International Institute of Physics, Universidade Federal do Rio Grande do Norte, 59078-400 Natal-RN, Brazil}

\author{K. Simonov}
\affiliation{Fakult\"{a}t f\"{u}r Physik, Universit\"{a}t Wien, Boltzmanngasse 5, 1090 Vienna, Austria}

\author{A.  Capolupo}
\affiliation{Dipartimento di Fisica ``E.R.Caianiello'', Universit\`{a} di Salerno,
and INFN Gruppo collegato di Salerno, Fisciano (SA) - 84084, Italy}

\author{B. C. Hiesmayr}
\affiliation{Fakult\"{a}t f\"{u}r Physik, Universit\"{a}t Wien, Boltzmanngasse 5, 1090 Vienna, Austria}


\begin{abstract}
Frustration of classical many-body systems can be used to distinguish ferromagnetic interactions from anti-ferromagnetic ones via the Toulouse conditions. A quantum version of the Toulouse conditions provides a similar classification based on the local ground states. We compute the global ground states for a family of models with Heisenberg-like interactions and analyse their behaviour with respect to frustration, entanglement and degeneracy. For that we develop analytical and numerical analysing tools capable to quantify the interplay between those three quantities. We find that the quantum Toulouse conditions provide a proper classification, however, refinements can be found. Our results show how the different local ground states affect the interplay and pave the way for further generalisation and possible applications to other quantum many-body systems.
\end{abstract}

\maketitle

\section{Introduction}

The manifestation of entanglement is a physical phenomenon which is completely absent in our everyday world. In recent years it has been shown that entanglement plays a fundamental role in the physics of quantum many-body systems~\cite{Amico2008}. Particularly, the violation of the entanglement area law~\cite{Holzhey1994,Vidal2003,Korepin2004,Calabrese2004} may serve as a detection tool of the presence of topologically ordered phases of matter~\cite{Chen2010}. Entanglement plays a crucial role also in local thermalization processes~\cite{Deutsch1991,Srednicki1994,Popescu2006,Barthel2008,Yang2017} and in the selection processes of specific states from degenerate ground state spaces~\cite{Hamma2016}. Moreover, the absence of entanglement for some ground states may provide a classification of the type of macroscopic phases~\cite{Giampaolofatt1,Giampaolofatt2,Giampaolofatt3}. The entanglement can be also included to the powerful matrix-product-state approach allowing for studies of entanglement-triggered processes~\cite{Murg2013,Barcza2015}.

Contrary to entanglement, frustration is a property that is present both in the classical and quantum domain.
It occurs when there are competing constraints, which cannot be fulfilled simultaneously~\cite{Toulouse1977-1,Toulouse1977-2,Lacroix2011,Diep2013}. In classical many-body physics, frustration plays a key role in the theory of the spin glasses~\cite{Villain1977,Kirkpatrick1977,Binder1986,Mezard1987}. Several quantum many-body systems are characterized by frustration whereas their classical counterparts are frustration-free~\cite{Giampaolo2011}. This suggests that frustration and entanglement are intimately related.

The aim of this paper is to characterize and understand the interplay between frustration and entanglement in a complex many-body system. For that we will take advantage of a recently introduced measure of frustration~\cite{Giampaolo2011,Marzolino2013}. Such a measure allows to extend the classical Toulouse conditions (CTCs)~\cite{Toulouse1977-1} to quantum regime, the quantum Toulouse conditions (QTCs)~\cite{Giampaolo2011}. The CTCs classify the interactions into two types, ferromagnetic and anti-ferromagnetic ones, depending on the local ground state space. A classical system is frustration-free if it can be transformed by unitary single-spin operations into a ferromagnetic one, else it is frustrated. In a similar manner the QTCs provide a classification into two types of interactions depending on the local ground states. Namely, the local ground state corresponds to a singlet state, a maximally entangled antisymmetric Bell state $\ket{\psi^{-}}=\frac{1}{\sqrt{2}}(\ket{\uparrow\downarrow}-\ket{\downarrow\uparrow})$, or to one of the three other maximally entangled Bell states (triplet states). Quantum systems with no geometrical frustration can be transformed by unitary and anti-unitary single spin operations into a model with local ground states formed by one of the triplet states.

This paper studies violations of the QTCs. For this purpose we have searched for a family of models which allow to distinguish between different sources of frustration -- geometrical and quantum ones -- via the QTCs. This rules out models with external fields~\cite{Barouch1970,Barouch1971}, cluster~\cite{Smacchia2011,Giampaolo2014,Giampaolo2015-1} or Dzyaloshinskii-Moriya terms~\cite{Dzyaloshinskii1958,Moriya1960}. We found that the anisotropic spin-$\frac{1}{2}$ Heisenberg models do the job. They are defined on a one-dimensional ring with translational symmetry established by an even number of spins. The strength of the interaction is chosen to be proportional to a power $\alpha \ge 0$ of the inverse distance between two spins. This distance-dependent interaction allows us to pass continuously between two extreme cases and herewith study the frustration and entaglement features under different conditions. This allows us to draw general conclusions on the interplay between them.

The paper is organized as follows. After the definition of the family of two-body Heisenberg-like Hamiltonians (Section~\ref{family}) we analyse the classical limit, which features frustration originated solely from the geometrical sources (Section~\ref{withoutentanglement}). These results serve as a benchmark for those of the models featuring frustration originated from quantum sources (Section~\ref{withentanglement}). In particular we show how the QTCs and their violations allow for a classification of these quantum models. In Section~\ref{conclusion} we summarize our results and discuss their relevance.

\section{Family of many-body systems}
\label{family}

In the following analysis we consider a system of dynamics governed by two-body Heisenberg-like interactions which feature frustration originated from easily distinguishable sources. For the sake of simplicity we consider a system defined on a one-dimensional ring of even number $N$ of spin-$\frac{1}{2}$ particles with periodic boundary conditions, and assume that it is invariant under spatial translation. Moreover, we assume that the strength of interaction depends on a power function of the inverse distance between two arbitrary spins. In this way, we are able to tune the weight of the quantum and geometrical sources of frustration presented in the system. The family of Hamiltonians reads
\begin{equation}
 \label{generalHamiltonian}
 \!\hat{H}\! = - \sum_{j>i} \frac{1}{d_{ij}^\alpha} \left( J_x\, \hat{\sigma}_i^x \hat{\sigma}_j^x+ J_y\, \hat{\sigma}_i^y \hat{\sigma}_j^y+J_z\,
 \hat{\sigma}_i^z \hat{\sigma}_j^z \right) \!,
\end{equation}
where $\hat{\sigma}_i^\mu$ with $\mu=x,\, y,\, z$ are the Pauli operators acting on the $i$-th spin, $J_\mu$ are
the anisotropy parameters in the spin direction $\mu$, which for the sake of simplicity we assume independent on the relative distance $d_{ij}=\min(j-i,i+N-j)$ between two spins.

The relative strength of the interactions is tuned by the parameter $\alpha \in [0,\infty)$. The models with $\alpha=0$ feature infinite-range interactions, namely each spin of the lattice interacts with the rest of spins with equal strength. On the contrary, the limit $\alpha\rightarrow\infty$ characterizes the models with short-range interactions. The in-between values of $\alpha$ establish a family of models with long-range interactions, i.e. each spin of the lattice interacts with the rest of spins with a strength that depends on the relative distance. For certain cases we have tested different dependences of the interactions on the distance, particularly an exponential dependence. However, no qualitative difference from the results presented in the paper was observed.

Without loss of generality we fix $J_z=-1$ and consider the behaviour of the system, by varying only the anisotropies $J_x$ and $J_y$ in the interval $ [-1 , 1] $. Under this assumption, we observe three regions in the $\{J_x, J_y\}$ space with different local ground states. The models of the region $J_y < -J_x$ feature the local ground state represented by the Bell state $\ket{\psi^{-}}=\frac{1}{\sqrt{2}}(\ket{\uparrow\downarrow}-\ket{\downarrow\uparrow})$, which is the singlet state. The models of the region $J_y > -J_x$ feature the local ground state represented by the Bell state $\ket{\psi^{+}}=\frac{1}{\sqrt{2}}(\ket{\uparrow\downarrow}+\ket{\downarrow\uparrow})$. Both these two states are maximally entangled and are equivalent with respect to local unitaries. This property plays a central role in the quantum information theory~\cite{Plenio2007} and is relevant in experimental realizations of mixed states~\cite{Carvacho2015}.

Models at the boundary between those two regions -- the case $J_y=-J_x$ -- provide a twofold degenerate local ground state, a convex combination of the Bell states $\ket{\psi^{-}}$ and $\ket{\psi^{+}}$, which represents a completely classical mixed state.

In the following analysis we focus mainly on three physical quantities,
\begin{itemize}
 \item the degeneracy $D(N)$ of the global ground state, which is fundamental for classical frustrated models~\cite{Kirkpatrick1977};
 \item the concurrence $C_{ij}(J_x,J_y,\alpha,N)$~\cite{Hill1997,Wootters1998}, which is a measure of the pairwise entanglement between two spins $\{i, j\}$ in the lattice;
 \item the measure of frustration $f_{ij}(J_x,J_y,\alpha,N)$ introduced in Ref.~\cite{Giampaolo2011,Marzolino2013}, which quantifies the distance between the local ground state and the state effectively realized by means of competing interactions.
\end{itemize}
We choose the concurrence as a valuable measure of entanglement for two reasons,
\begin{enumerate}[label=(\alph*)]
 \item the concurrence is one of the few measures of entanglement which is computable for mixed states, therefore it is also applicable for a system with a degenerate global ground state,
 \item there is a striking relation between frustration and concurrence if the local ground state is not degenerate~\cite{Giampaolo2015-2}, namely that the concurrence between directly interacting spins $C_{ij}(J_x,J_y,\alpha,N)$ is bounded by
\begin{equation}
 \label{concurrence}
 C_{ij}(J_x,J_y,\alpha,N) \ge \max\{0,1-2 f_{ij}(J_x,J_y,\alpha,N)\}.
\end{equation}
\end{enumerate}

Both the frustration and concurrence of a pair of spins $\{i,j\}$ can be calculated from the reduced density matrix $\hat{\rho}_{ij}(J_x,J_y,\alpha,N)$ obtained by tracing the maximally mixed ground state over the rest of spins of the lattice. Note that the reduced density matrix holds the same symmetry properties as the Hamiltonian since we consider maximally mixed ground states. In particular, the frustration of a directly interacting pair of spins $\{i, j\}$ is given by
\begin{eqnarray}\label{frusti1}
 f_{ij}(J_x,J_y ,\alpha,N) && \\
\nonumber && \!\!\!\!\!\!\!\!\!\!\!\!\!\!\! = 1-\mathrm{Tr}\Bigl(\hat{\Pi}(J_x,J_y) \cdot \hat{\rho}_{ij}(J_x,J_y,\alpha,N)\Bigr),
\end{eqnarray}
where $\hat{\Pi}(J_x,J_y)$ is the projector operator on the local ground space. It should be noted that the projector $\hat{\Pi}(J_x,J_y)$ does not depend on $i$, $j$ and $\alpha$.

In the case of classical models the reduced density matrix $\hat{\rho}_{ij}(J_x,J_y,\alpha,N)$ can be evaluated analytically. On the other hand, for the quantum models we use a numerical analysis based on the exact diagonalization of the Hamiltonian through the L\'{a}nczos algorithm. We do a complete numerical analysis for $N$ running from $6$ to $16$. For a limited number of models our analysis runs up to $N=20$. Our results allow to conclude that the features of the system can be described by finite $N$.

\section{The classical limit}\label{withoutentanglement}

We start our analysis of the frustration by considering classical models which provide a benchmark for the discussion of the quantum models. By choosing $J_x = J_y = 0$ we reduce the family of the models described by the Hamiltonian~(\ref{generalHamiltonian}) to an antiferromagnetic Ising model with long-range interactions, i.e.
\begin{equation}
 \label{classicalHamiltonian}
 \hat{H} = \sum_{j> i} \frac{1}{d_{ij}^\alpha}\; \hat{\sigma}_i^z \hat{\sigma}_j^z,
\end{equation}
where $\alpha$ is the parameter which tunes the relative strength of the interactions. The reduced Hamiltonian~(\ref{classicalHamiltonian}) consists of a sum of the terms proportional to $\hat{\sigma}_i^z \hat{\sigma}_j^z$ which commute with each other. Although this Hamiltonian is defined in the framework of quantum theory, it is equivalent to a classical model if the operators $\hat{\sigma}^z$ are replaced by a dichotomic variable restricted to the values $\pm 1$. In this case frustration can be originated only from geometrical sources.

The projector onto the local ground state for the models governed by the Hamiltonian~(\ref{classicalHamiltonian}) turns out to be a rang-two operator (see Sec.~\ref{family})
\begin{equation}
\nonumber\hat{\Pi}(0,0)= \ket{\uparrow \downarrow}\bra{\uparrow \downarrow}+\ket{\downarrow \uparrow}\bra{\downarrow \uparrow}\,.
\end{equation}
Taking into account the measure of frustration defined in Eq.~(\ref{frusti1}) and the invariance under global spin inversions of the Hamiltonian~(\ref{classicalHamiltonian}) we obtain the frustration of a pair of spins
\begin{equation}
\label{frusti1-classical}
f_{ij}(0,0,\alpha,N) = 1-\left(\rho_{ij}^{\uparrow \downarrow}(\alpha) + \rho_{ij}^{\downarrow \uparrow}(\alpha)\right)= 2\rho_{ij}^{\uparrow \uparrow}(\alpha),
\end{equation}
where $\rho_{ij}^{\eta\xi}(\alpha)$ is the matrix element of $\hat{\rho}_{ij}(J_x=0,J_y=0,\alpha,N)$ which corresponds to the projector $|\eta\xi\rangle\langle\eta\xi|$. In the following we discuss both the cases of finite-range interactions ($\alpha > 0$) and infinite-range interactions ($\alpha = 0$).

\subsection{Finite-range interactions ($\alpha>0$)}

\begin{figure}[t]
    \hspace{7cm}
    \includegraphics[width=8.8cm]{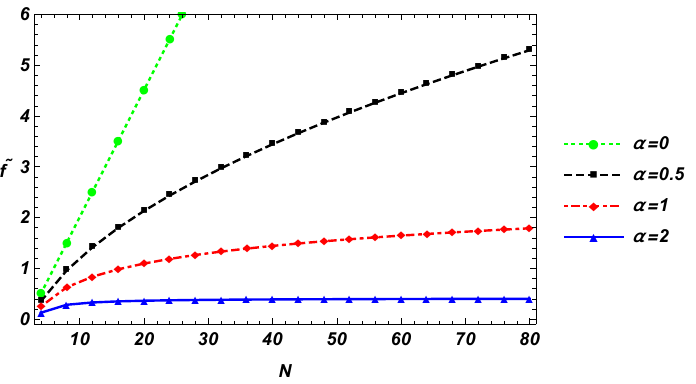}
     \caption{(Color online) Dependence of the characteristic frustration function $\tilde{f}$ for the classical models given in Eq.~(\ref{classicalHamiltonian}) on the number of spins $N$ for different values of $\alpha$. For $\alpha\leq 1$ the function $\tilde{f}$ diverges as $N$ increases, whereas for $\alpha > 1$ it converges to a finite value in the thermodynamic limit $N \rightarrow \infty$.
    }
    \label{figure1}
\end{figure}

For non-zero values of $\alpha$ the space composed by the ground states of the Hamiltonian~(\ref{classicalHamiltonian}) is a two-dimensional Hilbert space with a basis formed by two N\'{e}el states~\cite{Neel1948}. Such states are characterized by an alternation of spin up and spin down states.  Hence, the frustration vanishes for spins with an odd distance $d_{ij}$ and saturates to $1$ when the distance is even (see equation~(\ref{frusti1-classical})).

From a statistical point of view, the physical properties of the ground state (including its energy) are  affected essentially by the frustration of a strongly interacting pair of spins rather than weakly interacting ones. To capture this aspect we weight the frustration of each pair of spins $\{i,j\}$ by the relative strength of interaction between them and introduce a characteristic frustration function of the system of spins,
\begin{equation}
 \label{characteristicfrustration_general}
 \tilde{f}(J_x,J_y,\alpha,N)=\frac{1}{N}\sum_{j>i} \frac{f_{ij}(J_x,J_y,\alpha,N)}{d_{ij}^\alpha}\;.
\end{equation}
We point out that the characteristic frustration function~(\ref{characteristicfrustration_general}) is not a new measure of the frustration, but a weighted average value of the frustration presented in the system.

For the models described by the Hamiltonian~(\ref{classicalHamiltonian}) one finds analytical expressions of this characteristic frustration function $\tilde{f}(0,0,\alpha,N)$ for systems consisting of $N = 4k$ spins,
\begin{equation}
 \label{characteristicfrustration1}
 \tilde{f}(0,0,\alpha,N) = 2^{-\alpha} \cdot \sum_{i=1}^{N/4} \frac{1}{i^\alpha} - \frac{2^{\alpha-1}}{N^\alpha},
\end{equation}
and $N = 4k+2$ spins,
\begin{equation}
 \label{characteristicfrustration2}
 \tilde{f} (0,0,\alpha,N)= 2^{-\alpha} \cdot \sum_{i=1}^{(N-2)/4} \frac{1}{i^\alpha},
\end{equation}
where $k$ is an integer. Notice that the functions~(\ref{characteristicfrustration1}) and~(\ref{characteristicfrustration2}) coincide in the thermodynamic limit $N \rightarrow \infty$, indeed
\begin{equation}
 \label{characteristicfrustration3}
 \tilde{f}(0,0,\alpha,\infty) = \lim_{N\longrightarrow\infty} \tilde{f}(0,0,\alpha,N)=2^{-\alpha} \zeta(\alpha),
\end{equation}
where $\zeta(\alpha)$ is the Riemann zeta function, which converges to a finite value only for $\alpha > 1$, see Fig.~\ref{figure1}.


\subsection{Infinite-range interactions ($\alpha=0$)}\label{classical_infinite}

The case of the infinite-range interactions corresponds to the physical situation in which each spin is interacting with any other spin (equally in strength and independently of their distances). We find that the degeneracy of the global ground state of the system described by the Hamiltonian~(\ref{classicalHamiltonian}) depends now on the size of the system. In fact any state $|\uparrow^{N/2}\downarrow^{N/2}\rangle$ of $N/2$ spin up and $N/2$ spin down is a valid ground state of the system, thus the total degeneracy $D(N)$ of the ground state is
\begin{eqnarray}\label{equation12}
D_{\mathrm{classical}}(N)=\left(\begin{array}{c} N\\ \frac{N}{2}\end{array}\right) \simeq 2^N\sqrt{\frac{2}{\pi N}}
\end{eqnarray}
and hence increases exponentially with $N$.

For a maximally mixed state all the states with $N/2$ spin up and $N/2$ spin down have the same weight. The matrix element $\rho_{ij}^{\uparrow \uparrow}$ for a pair of spins has consequently to be proportional to the ratio of the number $\lambda(N)$ of states $|\uparrow^{N/2}\downarrow^{N/2}\rangle$ and the total degeneracy $D(N)$. The number $\lambda(N)$ can be obtained by fixing the pair of spins $\{i,j\}$ in the state $\ket{\uparrow\uparrow}$ and varying over all the possibilities for the remaining spins obeying the constrain of zero total magnetization,
\begin{eqnarray}\label{lambda}
\lambda(N) = \left(\begin{array}{c} N-2\\ \frac{N-4}{2}\end{array}\right).
\end{eqnarray}

After some algebra we obtain the frustration for each pair of spins in the lattice
\begin{eqnarray}\label{resultsep}
f_{ij}(0,0,0,N)= 2 \cdot \frac{\lambda(N)}{D_{\mathrm{classical}}(N)} = \frac{1}{2} \frac{N-2}{N-1},
\end{eqnarray}
which increases with the number of spins $N$ and converges to $\frac{1}{2}$ in the thermodynamic limit. Computing the characteristic frustration function~(\ref{characteristicfrustration_general}) we find
\begin{equation}\label{characteristicfrustration4}
 \tilde{f}(0,0,0,N)=\frac{1}{N} \cdot \frac{1}{2} \frac{N-2}{N-1} \cdot \frac{N (N-1)}{2} = \frac{N-2}{4}.
\end{equation}
Notice that the function~(\ref{characteristicfrustration4}) can be also obtained by performing the limit $\alpha \rightarrow 0^+$ of the characteristic frustration functions~(\ref{characteristicfrustration1}) and~(\ref{characteristicfrustration2}) for the lattices with finite-range interactions. In this way the characteristic frustration function $\tilde{f}$ is consistent and continuous even for the models with an exponential dependence of $D(N)$.

\section{Frustration originated from geometrical and quantum sources}\label{withentanglement}

In contrast to the models discussed in the previous section, the models with $J_y \neq -J_x$  have Hamiltonians which cannot be separated into locally commuting terms. The loss of commutativity between local terms implies that frustration can originate both from geometrical sources (as for commuting models) and from quantum sources. We start the analysis by discussing the two extreme cases separately.

\subsection{Short-range interactions ($\alpha \rightarrow \infty$)}

Since we focus on the systems of even number of spins, the models with short-range interactions satisfy the QTCs~\cite{Giampaolo2011} and, hence, frustration cannot originate from geometrical sources. Therefore, following QTCs there must exist a sequence of unitary operations which maps a model with a local ground state $\ket{\psi^{+}}$ into a model with local ground state $\ket{\psi^{-}}$.

For our models such a map is obtained by the rotations of $\pm\frac{\pi}{2}$ around the $z$ axis of all $N/2$ even spins (or alternatively odd). Particularly, such operations change the sign of the Pauli operators $\hat{\sigma}_x \rightarrow -\hat{\sigma}_x$ and $\hat{\sigma}_y \rightarrow -\hat{\sigma}_y$ of any even (odd) spin of the lattice. This implies that the sign of the anisotropies $J_x$ and $J_y$ changes, hence such a sequence can be considered as a bijective map between the models located above and below the symmetry axis $J_y=-J_x$. In the following we compare the models above and below the symmetry axis, line $J_y = -J_x$ (compare also with Fig.~\ref{InfRangeFrustration}).

Since the bijective map consists of local unitary operations only, it does not change the amount of entanglement and frustration. If no geometric frustration exists in the system, then there is a perfect symmetry between frustration and entanglement with respect to the line $J_y=-J_x$. This fact implies that models, which admit the singlet state $\ket{\psi^{-}}$ (above the line $J_y=-J_x$) and one of the components of the triplet state (below the line $J_y=-J_x$), behave in a similar way with respect to the entanglement and frustration. This symmetry is in a perfect agreement with the behaviour of the classical models. In fact, in the limit $\alpha \rightarrow \infty$ the classical models allow to pass from ferromagnetic to antiferromagnetic phase by means of local unitary operations only.

\begin{figure}[t]
    \hspace{7cm}
    \includegraphics[width=8.8cm]{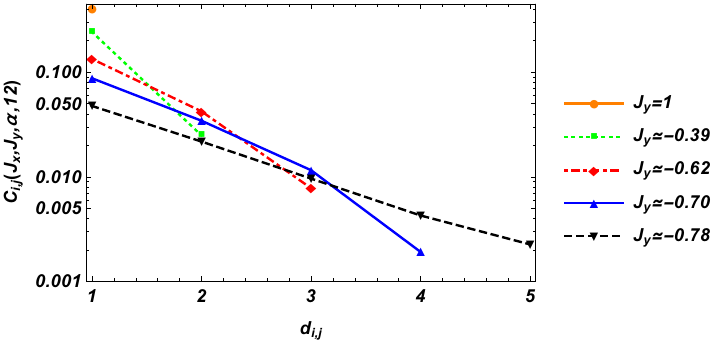}
     \caption{(Colour online) Behaviour of concurrence $C_{ij}(J_x, J_y, \alpha, N)$ of a pair of spins $\{ i, j\}$ as a function of the distance $d_{ij}$ between them for the several values of anisotropy $J_y$ with the fixed values of the anisotropies $J_x=0.9$ and $J_z=-1$. If $J_y$ approaches to $-J_x$ the amount of concurrence starts to increase. Numerical simulations are performed for a system of $N=12$ spins.
    }
    \label{figure2}
\end{figure}

Considering the entanglement properties of the system, it should be mentioned that the concurrence can hold a non-zero value even for a pair of non-interacting spins. As shown in Fig.~(\ref{figure2}) the amount of the concurrence tends to diverge if the set of the anisotropies approaches to the case $J_y=-J_x$. On the contrary, the value of each single concurrence decreases. This behaviour is in perfect agreement with the results of Ref.~\cite{Amico2006} which have studied the divergence of the amount of concurrence close to the factorization point for models with short-range interactions.

Summarizing, a perfect symmetry exists between models which have a singlet state or one of the triplet states as the local ground state, if frustration is originated from quantum sources only, i.e. no geometrical sources are present. In the infinite-range case such a symmetry breaks (as we show in the following).

\subsection{Infinite-range interactions ($\alpha = 0$)}\label{IRM}

\begin{figure}[t]
 \includegraphics[width=8.cm]{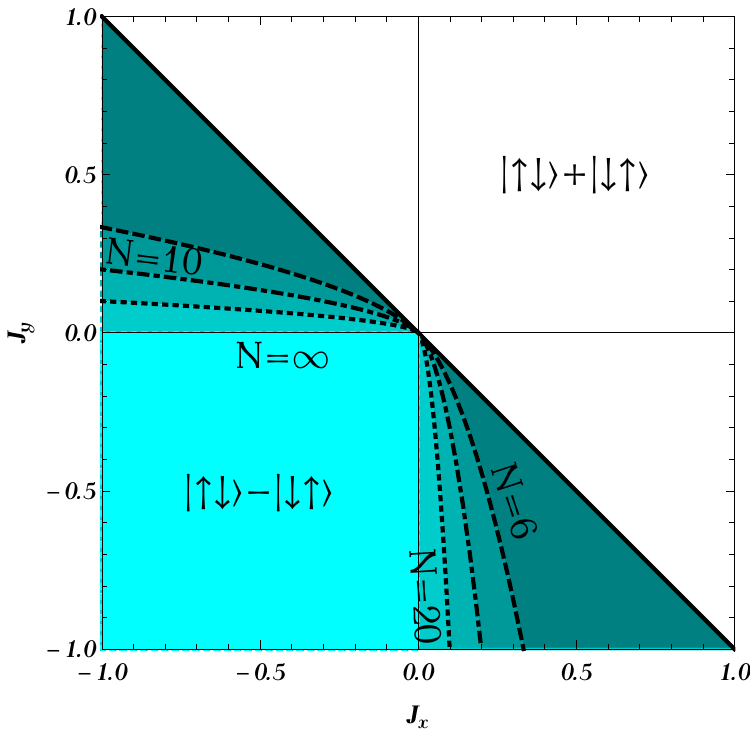}
 \caption{(Colour online) Classification scheme for the systems with $J_z=-1$ in dependence of $J_x,J_y$ for $\alpha=0$. For values above the line $J_x=-J_y$ (white region) the family of Hamiltonians admits a local symmetric ground state $\ket{\psi^+}=\frac{1}{\sqrt{2}}\{\ket{\uparrow\downarrow}+\ket{\downarrow\uparrow}\}$. On the contrary, below the line $J_x=-J_y$ (coloured region) the local ground states are still maximally entangled, but via the antisymmetric Bell state $\ket{\psi^-}=\frac{1}{\sqrt{2}}\{\ket{\uparrow\downarrow}-\ket{\downarrow\uparrow}\}$. Above the dotted curves, depending on the size $N$, lies the family $A$ of the models, which are characterised by non-degenerate global ground state. The remaning region features the family B of the models, which are characterised by degenerate global ground states. Particularly, in the thermodynamic limit the global ground state is degenerate only if both parameters $J_x,J_y$ are negative. The plots are generated by interpolating different numerical results for a given $N$  ($101\times 101$ models for $N\leq 14$ and $51\times 51$ models for $N>14$).}
 \label{InfRangeFrustration}
\end{figure}

In strong contrast to discussed short-range models one observes a strong dependence of the behaviour for values of $J_x,J_y$ above or below the symmetry line $J_y=-J_x$, which is illustrated in Fig.~(\ref{InfRangeFrustration}).

The models located above the line (white area in Fig.~(\ref{InfRangeFrustration})), i.e. those with $\ket{\psi^+}$ as a local ground state, are always described by the non-degenerate Hamiltonian. In this case, any single pair of spins has a value of the frustration in the interval $\Bigl[ \frac{1}{2} \frac{N-2}{N-1}, \frac{1}{2} \Bigr]$. Notice that $\frac{1}{2} \frac{N-2}{N-1}$ coincides with the value of frustration in Eq.~(\ref{resultsep}). The minimum of frustration is reached for models with $J_x=J_y>0$. In general, the numerical analysis shows that the value of frustration of each single pair of spins does not change along the line
\begin{eqnarray}
J_y(\gamma)&=& \gamma (J_x + 1) - 1 \qquad\forall\; 0\leq\gamma\leq\infty\;.
\end{eqnarray}
This unexpected identity can be explained by noting that the Hamiltonian~(\ref{generalHamiltonian}) commutes with the total magnetization $\hat{S}_z=\sum_i\hat{\sigma}_i^z$ along the $z$-axis if $J_x=J_y$. Therefore, the spectrum of the Hamiltonian can be decomposed into $N+1$ non-interacting samples as can be done in the classical limit as well. However, it should be noted that the degeneracy of the global ground state in the classical limit analysed in Sec.~\ref{classical_infinite} increases exponentially with $N$, whereas in the quantum models we have no degeneracy at all.

On the contrary, the limits $\gamma\rightarrow 0,\infty$ imply a maximum value of frustration of $\tilde{f}(J_x,J_y,0,N) = 1/2$. This value is independent on the size of the system, while the minimum value of frustration tends to $1/2$ in thermodynamic limit. Therefore, we conclude that the value of frustration is equal to $1/2$ in the whole region in thermodynamic limit. Hence, for a macroscopic system the amount of frustration in a model with infinite-range interactions which admits as local ground states $\ket{\psi^+}$ is indistinguishable from the ones of a classical model. Note that this indistinguishability appears though the degeneracy of the classical models is non-zero and can increase exponentially with $N$. This is a typical behaviour of frustration when classical and quantum models are compared~\cite{Moessner2001}.

Since all the interactions have the same strength, the concurrence is also independent on the distance between the spins. We find a saturation of the inequality~(\ref{concurrence}) by concurrence. Therefore, it is bounded from above by $\frac{1}{N-1}$ and approaches zero in the thermodynamic limit accordingly to the monogamy relation~\cite{Giampaolo2015-2}.

Turning to the models with $J_y<-J_x$, we obtain a completely different picture. This family of models falls into two different cases which we denote by $A$ and $B$. They are separated by the boundary which depends on the size of the lattice $N$, as shown in Fig.~\ref{InfRangeFrustration}. If the anisotropy parameters $J_x$ and $J_y$ of a model lie below the line $J_y=-J_x$ and above the lines
\begin{eqnarray}
\label{border}
 J_y(N) & = & -J_x -\frac14 N\left(N-2\right) J_x^2 \; \;  \mathrm{for} \; J_x \ge 0 , \\
 J_x(N) & = & -J_y -\frac14 N\left(N-2\right) J_y^2 \; \; \mathrm{for} \; J_x \le 0 ,  \nonumber
\end{eqnarray}
then it belongs to the family $A$, otherwise it belongs to the family $B$. Hence the number of models belonged to the family $A$ increases with the size of the lattice $N$.

For both the families the local ground states are unique, while the global ones differ. We observe that the models of the family $A$ can be considered as a continuation of those belonged to the region analysed before, except that the value of frustration approaches $1$. In this case the relation between frustration and concurrence~\cite{Giampaolo2015-2} is not saturated, and we have to evaluate the value of frustration directly from the reduced density matrix. Analysing the behaviour of the global ground state, we observe that the values of concurrence can be considered as an analytic continuation of the values obtained for the models with $J_y>-J_x$. Therefore, the family $A$ also features concurrence of each single pair which is bounded from above by $\frac{1}{N-1}$ and vanishes in the thermodynamic limit.

On the other hand, the family $B$ exhibits a degeneracy of the ground state which increases exponentially with $N$. Such a behaviour is similar to the one of the classical models with $J_x=J_y=0$ discussed in Section~\ref{withoutentanglement}. In more details, due to our numerical analysis the degree of degeneracy of the models in the family $B$ is equal to $\frac{1}{2+N}$ times the degeneracy~(\ref{equation12}) of the classical model,
\begin{eqnarray}\label{equation123}
D_{\mathrm{quantum}}(N)=\frac{D_{\mathrm{classical}}(N)}{2+N} \simeq 2^N \sqrt{\frac{2}{\pi N^3}}\;.
\end{eqnarray}
The value of frustration of each single pair of spins depends on the size of the lattice $N$ and is proportional to the ones obtained for the classical model,
\begin{eqnarray}
f_{ij}(J_x,J_y,0,N)=\frac{3}{4} \frac{N-2}{N-1}=\frac{3}{2} f_{ij}(0,0,0,N).
\end{eqnarray}
Consequently, the characteristic frustration function becomes
\begin{eqnarray}\label{equation18}
\tilde{f}(J_x,J_y,0,N)= \frac{3}{8} (N-2) \;.
\end{eqnarray}
It is important to outline that for the family $B$ as well as for the family $A$ the degeneracy of the ground state and the value of frustration are independent on the anisotropy parameters $J_x$ and $J_y$. Considering Eq.~(\ref{equation18}) and taking into account Eq.~(\ref{concurrence}) we may immediately conclude that for all the models in the region we have that the concurrence between two spins is always zero (independently on $N$). Such a behaviour of the degeneracy and frustration as well as the absence of entanglement make the models of family $B$ very close to those obtained by a classical limit (see the Section~\ref{withoutentanglement}).

\subsection{Long-range interactions ($0 < \alpha < \infty$)}

\begin{figure}[t]
\includegraphics[width=8.5cm]{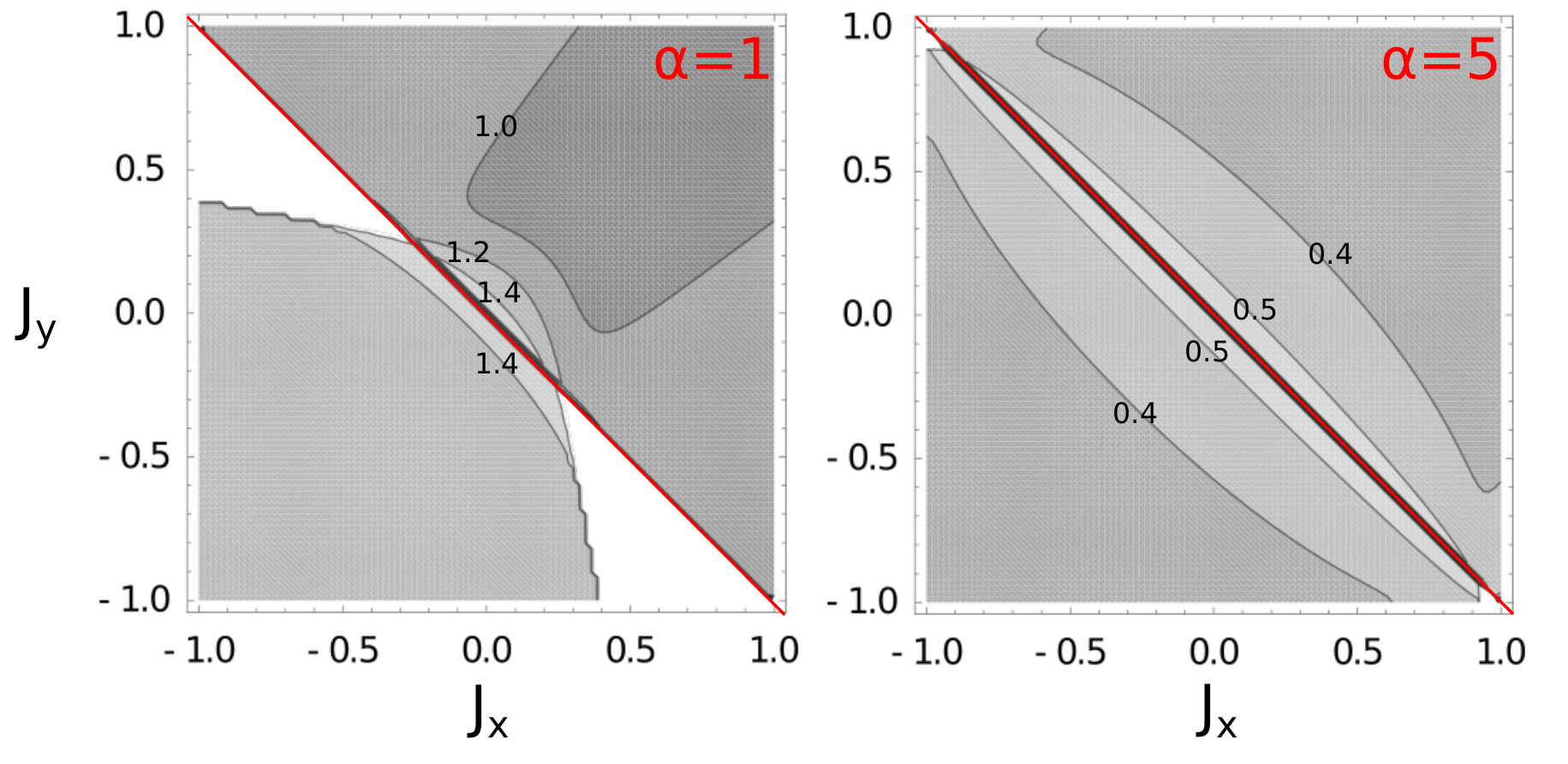}
\caption{(Colour online) Behaviour of the characteristic frustration function for $N=14$ spins for (a) $\alpha=1$ and (b) $\alpha=5$. Here lighter is the greyness, higher is the value of frustration. The white area corresponds to the models which feature the value of frustration equal to $1$. For the models with rather long-range interactions, $\alpha=1$, we observe a clear distinction between the families $A$ and $B$ of models below the line $J_x=-J_y$ as the one observed for the models with infinite-range interactions in Subsection~\ref{IRM}. With increasing $\alpha$ the family $A$ disappears almost completely. Furthermore, the difference in the behaviour of models with different local ground states disappears.}\label{figure3}
\end{figure}

The behaviour of the characteristic frustration function $\tilde{f}(J_x, J_y, \alpha, N)$ as a function of the anisotropy parameters for $\alpha=1,5$ are shown in  Fig.~(\ref{figure3}). For $\alpha=1$ we still observe the differences between the family of models A and B as discussed for $\alpha=0$. Note, however, that the macroscopic degeneracy observed for the family B for $\alpha=0$ disappears immediately for any $\alpha>0$. By increasing $\alpha$ the region of models of type A decreases and vanishes for $\alpha\rightarrow\infty$. 

\begin{figure}[t]
 \includegraphics[width=7.cm]{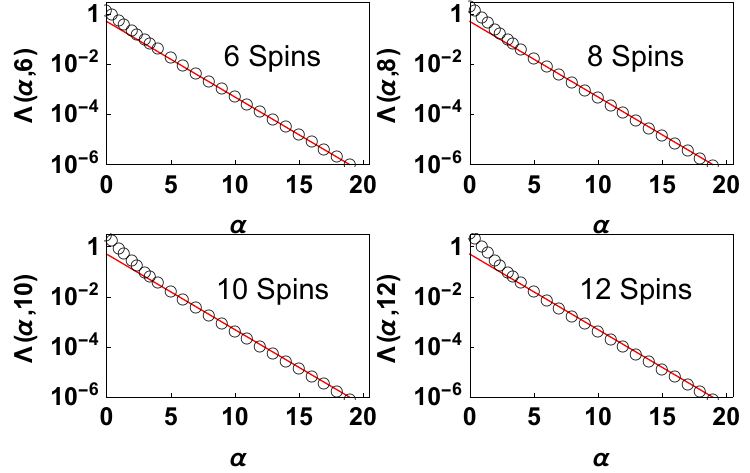}
 \caption{(Colour online) Behaviour of the asymmetry function $\Lambda(\alpha,N)$ scaling with $N=6,8,10$ in dependence on $\alpha$ (logarithmic scale). Each circle corresponds to the value of $\Lambda(\alpha,N)$ computed by integrating $10201$ models. The solid (red) lines show the exponential dependence of $\Lambda(\alpha,N)$, i.e. one proportional to $2^{-(\alpha+1)}$.}
 \label{figure5}
\end{figure}

To analyse this phenomenon quantitatively we introduce a new quantity sensitive to the difference between the frustration of the models, which lie in different subregions, namely above and below the line $J_y=-J_x$. Starting from $\tilde{f}(J_x, J_y, \alpha, N)$ we introduce as an anisotropy estimator between the symmetric and antisymmetric local ground state spaces the asymmetry function $\Lambda(\alpha,N)$ defined as
\begin{equation}\label{lambdadef}
\Lambda(\alpha,N)= \!\int_\Omega\! \left|\tilde{f}_{J_x,J_y}(\alpha,N)\!-\!\tilde{f}_{-J_x,-J_y}(\alpha,N)\right|d\Omega,
\end{equation}
where $\Omega$ is the parameter space of $\{J_x, J_y\}$. The behaviour of $\Lambda(\alpha,N)$ as a function of $\alpha$ for different $N$ is shown in Fig.~(\ref{figure5}), which exhibits its exponential decrease proportional to $2^{-(\alpha+1)}$ for $\alpha\geq 3$ independently of the size of lattice. Such a behaviour agrees with the results obtained for the two limiting cases. Indeed, in the limit $\alpha\rightarrow\infty$, which implies a symmetry between the models lied below and above the line $J_y=-J_x$, the asymmetry function vanishes for all $N$. In the opposite limit $\alpha \rightarrow 0$ we obtain an increase of the asymmetry function with $N$, which tends to $\Lambda(0,\infty) = \frac{3}{8}$ in the thermodynamic limit.

Since there is no finite critical value of $\alpha$ implying the full disappearance of family A, the characteristic frustration function can be bounded from above only for a given $\alpha$ and $N$, i.e. by
\begin{equation}
 \label{bound1}
 \tilde{f}(J_x,J_y,\alpha,N)\le 2\sum_{i=1}^{N/2}\frac{1}{i^\alpha}-\left(\frac{2}{N} \right)^\alpha\;,
\end{equation}
which becomes in thermodynamic limit
\begin{equation}
 \label{bound2}
 \lim_{N\rightarrow\infty}\tilde{f}(J_x,J_y,\alpha,N)\;\le\; 2 \zeta(\alpha)\;.
\end{equation}

\begin{figure}[t]
 \includegraphics[width=7.cm]{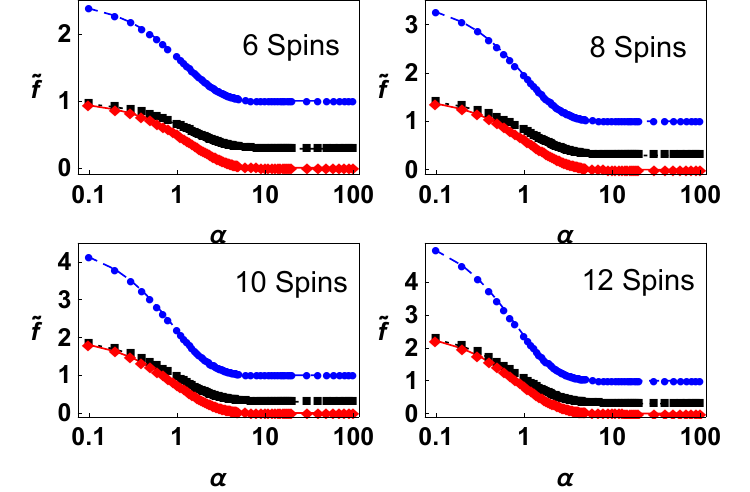}
 \caption{(Colour online) Behaviour of the characteristic frustration function $\tilde{f}(J_x, J_y, \alpha, N)$ in dependence on $\alpha$ for $N=6,8,10,12$. The maximum (blue circles) and the minimum (black squares) over all $J_x, J_y$ with non-degenerate local ground states of $\tilde{f}$ are plotted as well as $\tilde{f}(J_x, J_y, \alpha, N)$ for the antiferromagnetic long-range Ising model (red diamonds).}
 \label{figure4}
\end{figure}

For the lower bound we use the numerical results obtained by the exact diagonalization of the Hamiltonian. The behaviour of the maximum and minimum of $\tilde{f}(J_x, J_y, \alpha, N)$ as a function of $\alpha$ for all the values of $J_x$ and $J_y$ with the fixed $N=6,8,10,12$ is shown in Fig.~(\ref{figure4}). While the maximum is given by Eq.~(\ref{bound1}), the minimum tends to coincide with the classical value for $\alpha\rightarrow0$ and remains non-zero in the limit $\alpha\rightarrow \infty$. Defining the minimum of the characteristic frustration function as
\begin{equation}
 \label{fbar}
\bar{f}(\alpha,N):=\min_{J_x,J_y}\tilde{f}(J_x,J_y,\alpha,N),
\end{equation}
we find that it is bounded in the thermodynamic limit by
\begin{equation}
 \label{bound3}
2^{-\alpha} \zeta(\alpha) \le \bar{f}(\alpha,\infty) \le 2^{-\alpha} \zeta(\alpha) + \bar{f}(\infty,\infty),
\end{equation}
where $\bar{f}(\infty,\infty)$ is the minimum of the characteristic frustration function in the thermodynamic limit for the models with short-range interactions. It is a finite number, and $\zeta(\alpha)$ diverges for $\alpha\ge 1$, therefore, $\bar{f}(\alpha,\infty)$ diverges for $\alpha\ge 1$. Since the maximum of the characteristic frustration function diverges for $\alpha\ge1$ as well due to Eq.~(\ref{bound2}), we find that the models with long-range interactions ($\alpha \le 1$) can be divided into two families. The first family consists of models which imply a divergence of $\tilde{f}(J_x,J_y,\alpha,N)$ independently on the anisotropies $\alpha\leq 1$. The second family of models, $\alpha\geq 1$, exhibits a finite value of the characteristic frustration function even in the thermodynamic limit.

Notice that that beyond the limit $ \alpha = 1 $ the non-interacting fermionic models violate the area law of the entanglement entropy~\cite{Vidal2003,Vodola2016}. Hence, this result seems to suggest the existence of a connection between frustration and entanglement even deeper than the one known until now. Unfortunately, with our approach it cannot be analysed deeper.

The concurrence shows a behaviour similar to the one of frustration. For $\alpha>1$ the concurrence is limited to the next-neighbour pairs of spins with the exception of the models with $J_x$ and $J_y$ close to the factorization point. The value of concurrence depends on the choice of the anisotropy parameters, and the symmetry between the models with respect to the line $J_x=-J_y$ is recovered in the limit $\alpha\rightarrow\infty$. On the contrary, for $\alpha<1$ the range of concurrence diverges for the models which admit the singlet state $\ket{\psi^-}$ as a local ground state. Such a range is limited to the next-neighbour pairs of spins for the models which admit $\ket{\psi^+}$ as a local ground state. This fact proves, that a strong dependence of the properties of frustration on the local ground space highlights a peculiar behaviour of the properties of entanglement in the system.

\section{Summary and conclusions}\label{conclusion}

The aim of this contribution is to study how the different sources of frustration affect the properties of the global ground state of a many-body system and its interplay with entanglement. In particular, we have analysed a family of spin-$\frac{1}{2}$ Heisenberg-like Hamiltonians with distance-dependent two-body interactions and studied their properties related to frustration and entanglement. Since only few models allow to find analytical solutions, we have performed numerical computations for finite-size systems. The obtained results turn out to be very consistent with increasing number of spins. This fact allows for an extrapolation of behaviour of the system in the thermodynamic limit.

The presence or absence of geometrical frustration introduces a classification for the family of models which is completely equivalent to the ferromagnetic/antiferromagnetic classification via the classical Toulouse conditions (CTCs). The role played by the antiferromagnetic models in the classical case passes in the quantum regime to the models which admit the singlet state as the local ground state. Such models, in the case of infinite-range interactions, imply different physical properties of the global system with respect to models admitting one of the triplet states as the local ground state. In the short-range limit such a difference vanishes fully. We quantify this different physical behaviour by introducing an asymmetry function $\Lambda(\alpha,N)$. It characterizes the difference of frustration based on singlet or triplet states as local ground states, which are all four maximally entangled states differing only by local unitaries. By analysing the function $\Lambda(\alpha,N)$ we have proven quantitatively its exponential decrease depending on the range of the interactions.

Our results illustrate the role played by the different local maximally entangled ground states in physics of frustrated quantum systems. These results and the developed methods pave the way e.g. for a generalization of spin glasses~\cite{Binder1986} to the quantum regime~\cite{Sachdev1994}. Furthermore, our results suggest that an even deeper relation between entanglement and frustration exists.

\vspace{0.5cm}
\textbf{Acknowledgements:}
S.M.G. acknowledges financial support from the Ministry of Science, Technology and Innovation of Brazil, K.S. acknowledges the Austrian Science
Fund (FWF-P26783), A.C. acknowledges partial financial support from MIUR and INFN and B.C.H. Austrian Science Fund (FWF-P23627).

\end{document}